\documentclass[%
 sor,
 jor,
 amsmath,amssymb,
 reprint,%
]{revtex4-2}

\usepackage{graphicx}
\usepackage{dcolumn}
\usepackage{bm}

\begin{document}

\preprint{AIP/123-QED}

\title[]{Anti-hairpin vortices in the buoyancy-driven turbulent boundary layer}

\author{C. Metin and M. A. Ezan}
 \email{cagri.metin.ege@gmail.com}
 \email{mehmet.ezan@deu.edu.tr}
\affiliation{
Faculty of Engineering, Department of Mechanical Engineering, Dokuz Eylul University, Buca, Izmir, Turkey
}%


\begin{abstract}
Here we show the hairpin vortices point the upstream direction of flow instead of downstream, and this characteristic is found intrinsic behavior for turbulent buoyancy-driven boundary layer. We uncover the coherent vortices straddle high-speed streak instead of low-speed streak; moreover, the quadrant of the tangential Reynolds shear stresses dominates the first and third quadrant instead of the second and fourth one. These findings may help to improve our understanding of buoyancy-driven flow and buoyancy effects in turbulent models.
\end{abstract}


\maketitle

\section{\label{sec:level1}Introduction}

Buoyancy-driven flows are ubiquitous in nature and many engineering applications such as nuclear reactors \citep{chung2004thermal, lee2007natural}, or electronic industry \citep{incropera1988convection}. Moreover, many species in nature use updraughts to sustained static soaring for energy-efficient flight and choosing a particular flight path during seasonal migration \citep{shepard2016moving}. From this perspective, this study focuses on vortex dynamics to shed light on the mechanism of turbulent natural convection that may help to improve the efficiency of those devices and increase our knowledge of nature.

Strong turbulent natural convection in an enclosed cavity has been receiving increasing attention over the last decades \citep{miroshnichenko2018turbulent}; historically, the natural convection of the cavity has been studying for decades to now. Most of the former studies performed in the laminar regime \citep{de1983natural, markatos1984laminar, inaba1984natural} or weakly turbulent \citep{chenoweth1986natural, braga1992transient, tian2000low1, tian2000low2, zhang2014numerical} due to limited computational resources; conversely, the turbulent natural convection in the cavity had relied on experimental studies \citep{inaba1984natural, ivey1984experiments, lankford1986natural, schopf1995natural}. With increasing computational resources, researchers have been started to pay attention to strong turbulent flow at high Rayleigh numbers \citep{versteegh1998turbulent, fedorovich2009turbulent, barhaghi2007natural, peng2001large, kizildag2014large, kogawa2016large, ng2017changes, sebilleau2018direct}.

Coherent turbulent structures of the natural convection boundary layer have been investigated in a limited number of studies and need to be further research. More specifically, is the theory of the hairpin vortices in the canonical turbulent boundary layer is valid for the buoyancy-driven turbulent boundary layer? For comparison, the streaky flow was observed in the buoyancy-driven boundary layer; further, the streaks become finer as Rayleigh number increases \citep{ng2017changes}. The spanwise spacing of streaks was found 100-200 viscous wall units which are similar to the canonical turbulent boundary layer \citep{ng2017changes}. We should remind that typical spanwise spacing of hairpin vortex is about 100 viscous wall units and rides on the low-speed streaks \citep{adrian2007hairpin}. However, the existence of streaks is not mean the existence of hairpin vortices \citep{lozano2014time, hack2018coherent}. On the other hand, transitional spanwise vortices in the buoyancy-driven boundary layer eventually stretched out and generated the hairpin vortices \citep{barhaghi2007natural}. These coherent structures were also associated with the hot thermal plume from the hot wall \citep{sebilleau2018direct}; moreover, conditional averaged structures were related with a large temperature gradient fluctuation \citep{pallares2010turbulent}. The inclined angle of these vortices was about 30\textsuperscript{o} \citep{pallares2010turbulent} which is similar to the hairpin-like vortices in the canonical turbulent boundary layer. However, some open questions have still existed, more abstract, can we observe the same mechanism: generation, growth, and interaction of structures proposed in the hairpin vortex theory, or has the buoyancy-driven turbulent boundary layer intrinsic mechanism?

The purpose of this study is to examine the coherent turbulent structures in the buoyancy-driven turbulent boundary layer in a differentially heated cavity. In this approach, we conducted a direct numerical simulation at Rayleigh number 1.58 $\times$ 10\textsuperscript{9} for air. In particular, the emphasis is placed on the examination of how hairpin vortices relate to each other, and how and why they evolve in time \citep{lozano2012three} in the buoyancy-driven turbulent boundary layer. This point of view relies on the quasi-deterministic approach instead of the statistical events of the turbulent quantities (see \cite{robinson1991coherent}). In this approach, we focus on the kinematic description of vortical structures as well as the elucidation of how they related to each others.

\section{\label{sec:level2}Background: theory of hairpin vortex}

\begin{figure}
		\centerline{\includegraphics[width=\textwidth]{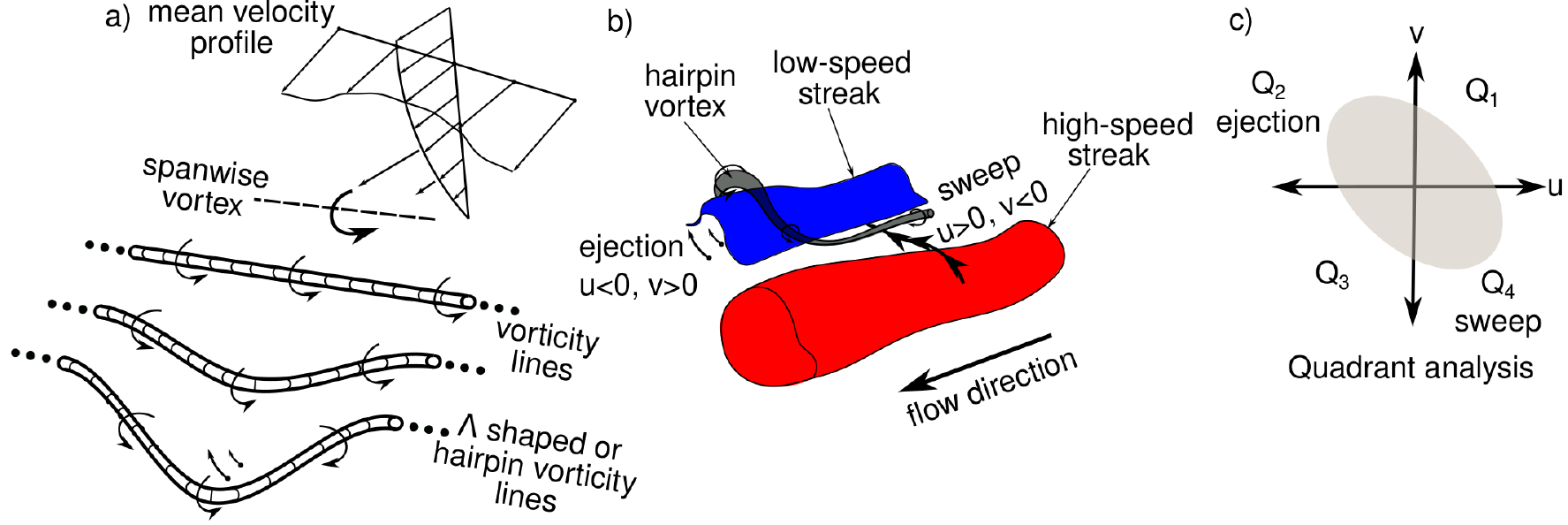}}
		
		\caption{Schematic of hairpin vortex theory. (\textit{a}) Conceptual illustration of the vortex filaments during the evolution of a hairpin vortex. (\textit{b}) The low-speed streak and high-speed streak with a hairpin vortex. (\textit{c}) Tangential velocity fluctuation (u in the streamwise direction and v is the wall-normal direction of velocity component) is mapped on the two-dimensional coordinate systems.}
		
		\label{fig:1}
	\end{figure}

	Even though the turbulent flow is treated complex, chaotic, and randomness motion, some coherent structures can be observed long enough with various scale. These structures are not only basic building blocks of turbulent flow but also significantly contribute to shear force and heat transfer \citep{adrian2007hairpin, robinson1991coherent}. A coherent turbulent structure in the boundary layer is demonstrated by spanwise oriented vortex lines in a boundary layer (see Fig. \ref{fig:1}a). This structure stretches and lifts away from the wall. The pattern of the vortex is identified as $\Lambda$ or hairpin shaped. The hairpin vortex straddles the low-speed streak associated with quasi-streamwise vortices shown in Fig. \ref{fig:1}b  \citep{adrian2007hairpin}. The spanwise-oriented head of the hairpin-like vortex violently lifts away from the wall, encounter larger mean flow velocity and retards the background motion (u $<$ 0 and v $>$ 0, where u and v velocity fluctuations in the streamwise and wall-normal direction, respectively) is also called ejection event \citep{adrian2007hairpin}. Meanwhile, the ejection event feeds by legs and necks of the hairpin vortex from the high-speed streaky flow; fluid moves toward the wall referred to sweep event (u $>$ 0 and v $<$ 0) \citep{adrian2007hairpin}. As basic building blocks of the turbulent boundary layer are hairpin vortices, the streamwise (u) and wall-normal (v) velocity fluctuation are correlated with each other, and more probability of u times v is negative (Fig. \ref{fig:1}c) \citep{adrian2007hairpin}. Quadrant analysis is usually addressed to explain the role of the Reynolds stresses on the ejection and sweep events. The ejection events (the second quadrant of tangential Reynolds stresses) are generally observed stronger and higher speed than the sweep events (the fourth quadrant of tangential Reynolds stress) \citep{adrian2007hairpin} meanwhile high-speed streaks generally tend to be larger than the low-speed streaks \citep{jimenez2018coherent}.
		
\section{\label{sec:level3}Organization of coherent vortical structures}

\begin{figure}
		\centerline{\includegraphics[width=\textwidth]{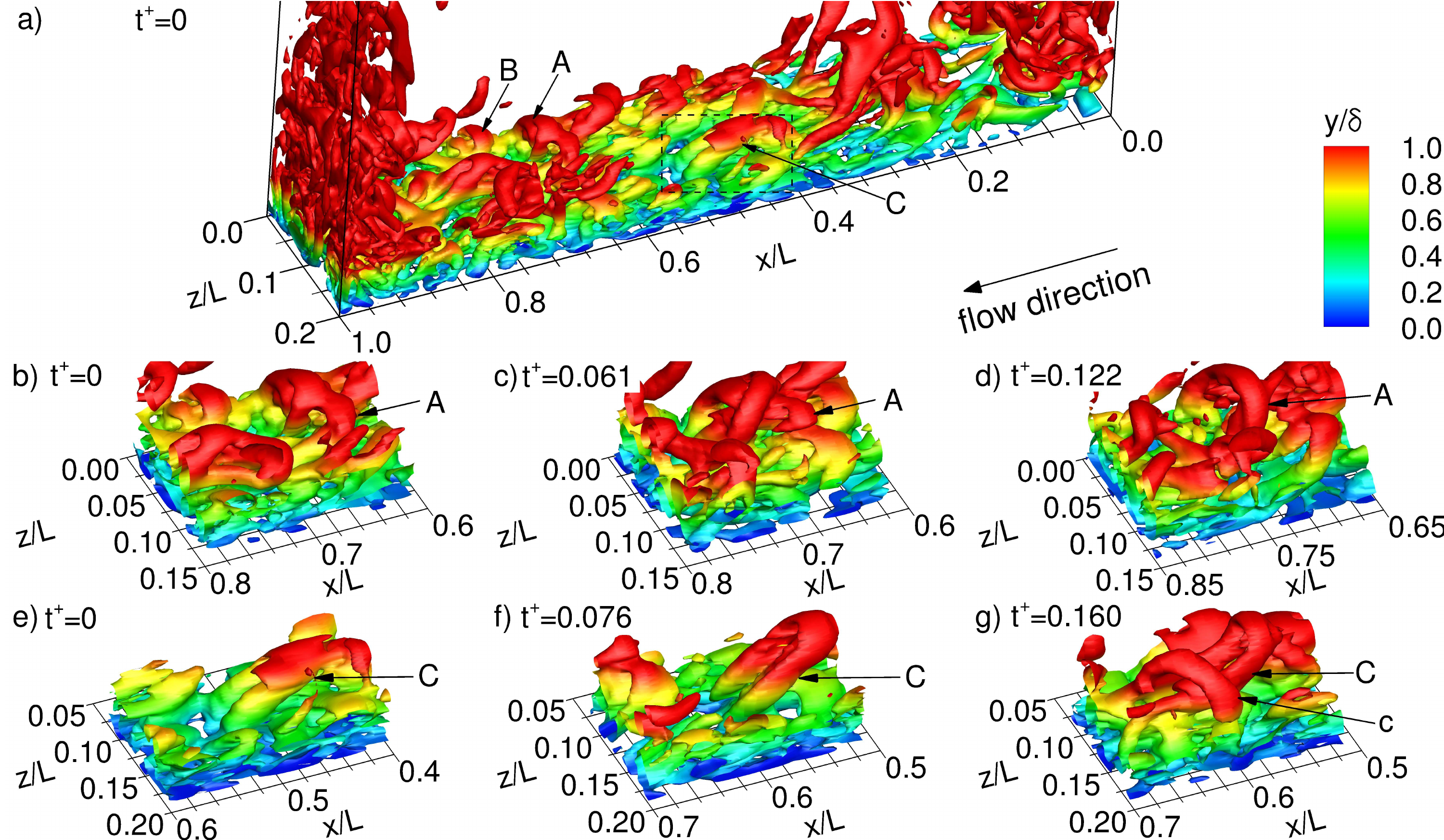}}
		
		\caption{Formation of anti-hairpin vortices in the buoyancy-driven turbulent boundary layer on the hot surface of the cavity. The vortical structures are identified by a surface of the swirling strength, $\lambda_{ci}$, with 5.55$\%$. The vortex core surface is colored the rainbow color scheme, and the color scale is varied according to the $y/\delta$, where $\delta$ is the boundary layer thickness at the middle of the cavity height. (\textit{a}) Overview of the several anti-hairpin vortices at $t^+$ = 0, which represents reference time value. (\textit{b}) The pre-merging state of the two anti-hairpin vortices labeled A. (\textit{c}) The annihilation of the closed legs of the anti-hairpin vortices during the merging process. (\textit{d}) The post-merging state of the two anti-hairpin vortices generated the larger vortex labeled A. (\textit{e}) A young anti-hairpin vortex generated via turbulent bulge. (\textit{f}) The inclined structure tends to about 45$^o$ to the wall. (\textit{g}) The regeneration of another young anti-hairpin vortex, labeled c, between the legs of the parent vortex, labeled C.}
		
		\label{fig:2}
	\end{figure}

The hairpin-like vortices are still dominant structures in the buoyancy-driven turbulent boundary layer and have similar features of the theory of hairpin vortex they nevertheless present distinguishable characters (Fig. \ref{fig:2}a). Merging mechanism of hairpin vortices could be many different configurations in the real turbulent boundary layer; however, three idealized possible merging mechanisms were presented by \cite{tomkins2003spanwise}. One of the proposed merging mechanism is the inner legs of the vortices join together in the spanwise direction, and the vortices growth with the annihilation of the inner legs and convect to one single larger vortex \citep{tomkins2003spanwise, wark1991experimental}. As shown in those vortices labeled A in Fig. \ref{fig:2}b, two hairpin vortices interacted in the spanwise direction. While dimensionless time differences reached t$^+$ = 0.061 (Fig. \ref{fig:2}c), the inner legs cannot be identified, and the heads of the vortices came to the verge of the merge each other. At the final stage of the spanwise merging (Fig. \ref{fig:2}d), a larger hairpin vortex was formed with the annihilation of neighboring legs as proposed \citep{tomkins2003spanwise}. However, the process is not symmetric and identical. The selenoid effects, analogous magnetic flux induced by windings, can be observed in vortices labeled B (Fig. \ref{fig:2}a). In this model, the conceptual scenario of the nested packets of the hairpin vortices aligns in the streamwise direction and induced by combined vorticity core \citep{adrian2000vortex}. Another common feature is that the head of the quasi-streamwise vortex pair lifts away from the wall (Fig. \ref{fig:2}e); intermediate structures inclined about 45 degrees to the wall afterward (Fig. \ref{fig:2}f). The subsequent structure is vortex labeled C, tends to the curved vertical direction to the wall \citep{zhou1999mechanisms} while moving downstream as well as becomes a mature hairpin vortex (Fig. \ref{fig:2}g). By t$^+$ = 0.160, the legs and necks of the hairpin vortex have a shallow angle to the wall while the head has a steeper inclination \citep{zhou1999mechanisms}. The parent-offspring mechanism (or regeneration mechanism) of the hairpin vortex was also observed in Fig. 2g. The process of replicate itself presented younger secondary hairpin vortex labeled c that straddled on the legs of the parent vortex labeled C at t$^+$ = 0.160 \citep{zhou1999mechanisms, panton2001overview}.

The hairpin vortices in the buoyancy-driven boundary layer have distinguishable differences than the theory of hairpin vortex. The most striking feature of these hairpin vortices is point upstream direction instead of the downstream (Fig. \ref{fig:2}a). Nevertheless, an upstream pointing of the hairpin vortex is observed in some previous studies, but this is not a common feature.  One of the cases observed upstream pointed hairpin represented transitional flow induced by passing wakes \citep{wu1999simulation}.  A large-amplitude vertical velocity component in the free stream velocity triggered the transition flow that produced upstream pointed turbulent spots. The main motivation of this behavior was related to the source of the turbulence. It was suggested that the upstream pointed spot was triggered a turbulence source higher above the wall; otherwise, the spot was pointed downstream direction. In the case of the varicose-like breakdown (or symmetric breakdown), the hairpin vortex alternatively pointed upstream or downstream direction \citep{wu1999simulation}. It was also observed downstream pointed hairpin vortices located above the legs of the upstream pointed hairpin vortex. It was suggested that interaction between high- and low-speed streaks play an important role in the formation of the hairpin vortex \citep{wu1999simulation}. However, in both case could not explain anti-hairpin vortex, they had neither large-amplitude vertical velocity nor varicose-like breakdown triggered. Moreover, the upstream and downstream pointed vortex was not observed together in this study.  Nonetheless, the case of the wall jet also produced an upstream pointed hairpin vortex, furthermore, all of the vortices pointed upstream in this case. Despite the other studies, one striking similarity between the hairpin vortex from wall jet and buoyancy-driven boundary layer flows is that the most probability of velocity fluctuation in the quadrant analysis was first and third quadrant \citep{nishino2010large}. However, some unanswered questions have still existed about the source of these vortices, and it was also stated that "the exact mechanism of the formation of hairpin vortices is still unclear" \citep{nishino2010large}. It is stated that the possible mechanism of the upstream pointed vortices was suggested centrifugal instability on a convex surface. Nevertheless, centrifugal instability is not a dominant phenomenon in the case of the buoyancy-driven boundary layer.

	\begin{figure}
		\centerline{\includegraphics[width=\textwidth]{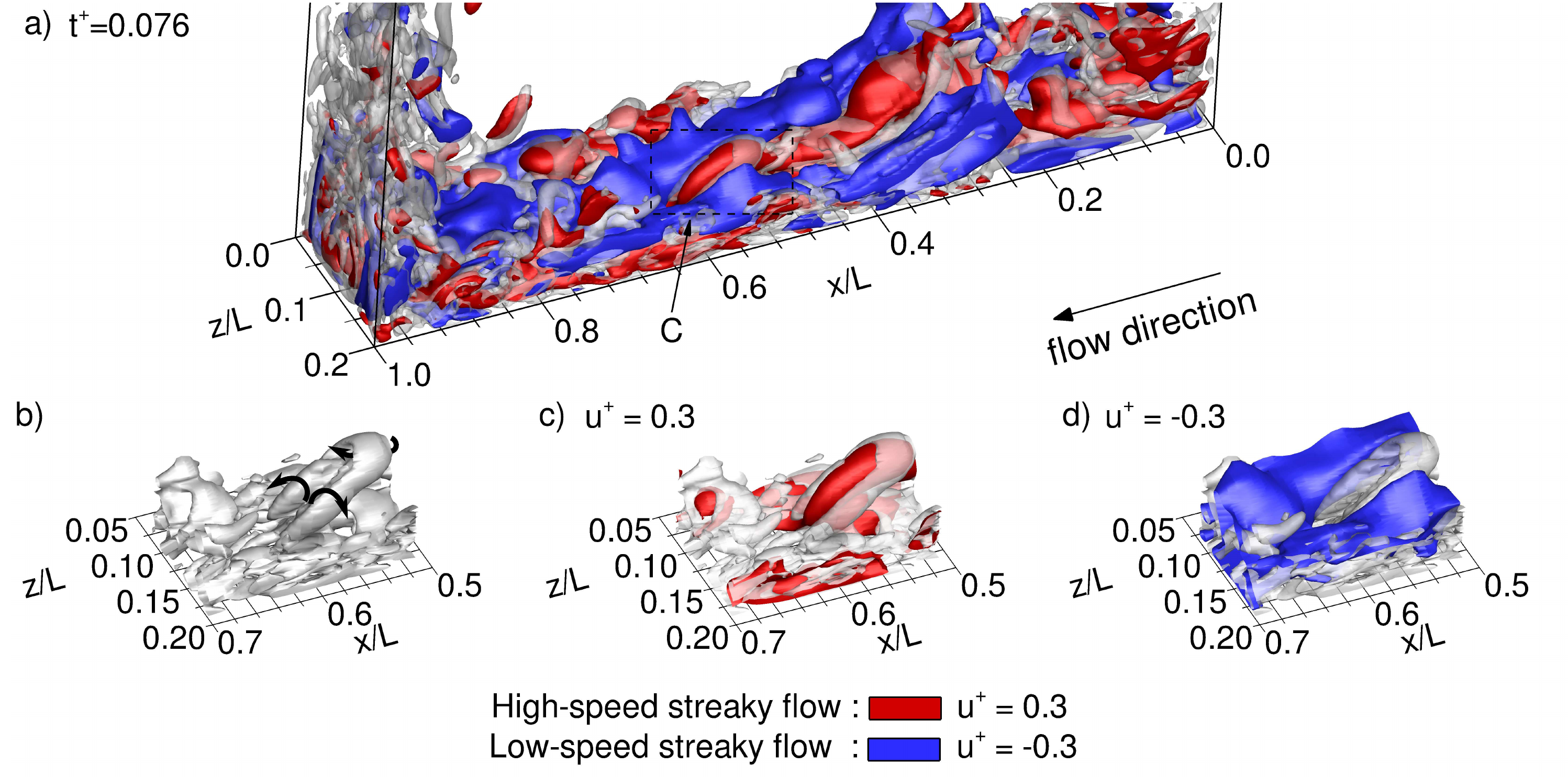}}
		
		\caption{High- and low-speed streaks with the anti-hairpin vortices. (\textit{a}) The high- and low-speed flow region with the anti-hairpin vortex at the $t^+$ = 0.076. (\textit{b}) The focused view on the labeled C vortex colored by white. The rotational direction of the legs and head of the vortex are represented by black arrows. (\textit{c}) The red-colored surface represents high-speed streaks with the transparent white-colored anti-hairpin vortex. (\textit{c}) The blue-colored surface represents low-speed streaks with the transparent white-colored anti-hairpin vortex.}
		
		\label{fig:4}
	\end{figure}

 One of the well-known characteristics of the hairpin vortex is streaky flow. The long and narrow elongated region of streamwise velocity fluctuation (u) in the turbulent flow is defined streaks and observed all distance from the wall in the turbulent boundary layer \citep{jimenez2018coherent}. Low-speed streaks  (u $<$ 0) generally associate with streamwise oriented vortex pair, and the head of the vortex lifts away on the low-speed streak. Similarly, the buoyancy-driven turbulent flow contains the streaks and accommodates several vortices on them (Fig. \ref{fig:4}a). Low-speed streaks play an important role in the skin friction, transport of turbulent kinetic energy, and vortex generation for the canonical turbulent boundary layer flow \citep{adrian2007hairpin}.  In the case of the anti-hairpin vortices, the high-speed streak is a source of heat transfer, skin friction, turbulent kinetic energy productions and dissipation, and the high-speed streaks also accommodate the several vortices. As shown in Fig. \ref{fig:4}b, labeled C vortex lifted away from the wall as proposed the hairpin theory; however, the vortex straddled on the high-speed streak instead of the low-speed streak (Fig. \ref{fig:4}c), and low-speed streak engulfed the high-speed streak and the anti-hairpin vortex (Fig. \ref{fig:4}d). The low-speed streaks generally explained that a hairpin vortex lifts from the wall as slower fluid particle upwash that retards the background flows \citep{adrian2007hairpin}. Whereas, the anti-hairpin vortex lifts away from the wall on the high-speed streaks, as well as, the head of the anti-hairpin vortex pushes the fluid particles to the downstream direction and accelerates it because of the orientation of the vortex (Fig. \ref{fig:4}b). Lastly, the high-speed streak generally tends to be larger than the low-speed streak in the canonical turbulent boundary layer \citep{jimenez2018coherent}, yet low-speed streak is observed larger than the high-speed streak in the buoyancy-driven turbulent boundary layer (Fig. \ref{fig:4}a).
 
 	\begin{figure}
		\centerline{\includegraphics[width=\textwidth]{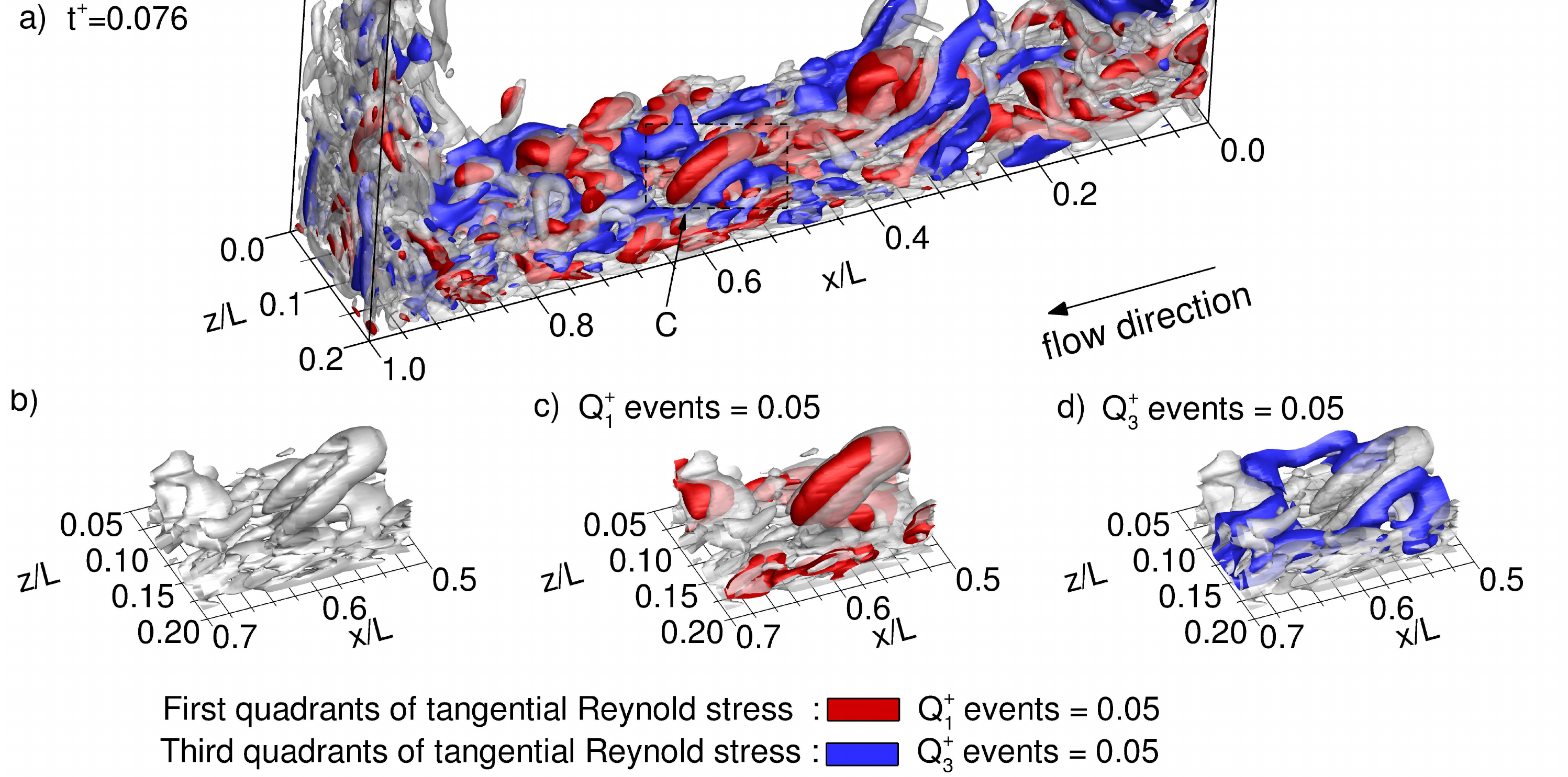}}
		
		\caption{First and third quadrant of the tangential Reynolds stresses with the anti-hairpin vortices. (\textit{a}) The first and third quadrant with the anti-hairpin vortex at the $t^+$ = 0.076. (\textit{b}) The focused view on the labeled C vortex colored by white. (\textit{c}) The red-colored surface represents the first quadrant of Reynolds stresses with the transparent white-colored anti-hairpin vortex. (\textit{d}) The blue colored surface represents the third quadrant of Reynolds stresses with the transparent white-colored anti-hairpin vortex.}
		
		\label{fig:5}
	\end{figure}
	
The early single-point experimental studies of hairpin vortices were inspected tangential Reynolds stresses in a wall-bounded turbulent boundary layer that was observed more probability of the second quadrant (related to the ejection event) and fourth quadrant (related to the sweep event) of the u-v plane, whereas these events are nowadays presenting three-dimensional structures in the modern computer era; these dominant events are the main source of the streamwise momentum exchange between the different layer of flow and turbulent drag \citep{jimenez2018coherent}. Unlike the behavior of tangential Reynolds stresses for the canonical turbulent boundary layer, the most dominant u-v correlation for the anti-hairpin vortices in the buoyancy-driven turbulent boundary layer are first and third quadrant of the u-v plane (Fig. \ref{fig:5}a) while the second and fourth quadrant of the tangential Reynolds stresses were found uncorrelated and unrelated to anti-hairpin vortices. In the case of anti-hairpin vortex, the ejection event corresponds to the first quadrant of the Reynolds stresses instead of the second one, which carries fluid particles away from the wall (Fig \ref{fig:5}c), whilst the sweep event corresponds to the third quadrant instead of the fourth one, that pushes the fluids toward to the wall (Fig \ref{fig:5}d). Notwithstanding, the ejection and sweep events correspond to the different quadrants than the canonical turbulent boundary layer, the three-dimensional ejection event tends to more larger than the sweep event in both cases \citep{jimenez2018coherent} that might be related to the ejection events supported by more extended duration sweep events, and the ejections tend to form in a group \citep{adrian2007hairpin}. Also, the ejection events focus in the concentrated region underneath of the head of the vortex as the sweep events prevail distributed region outboard of the vortex \citep{adrian2007hairpin}. It is stated that the vortices take the energy from the shear with u times v $<$ 0 and lose the energy with u times v $>$ 0 in the canonical turbulent boundary layer \citep{jimenez2018coherent}. Because of the most probability of correlated velocity component u times v $>$ 0 (positive quadrants) instead of u times v $<$ 0 (negative quadrant) in the anti-hairpin vortex theory, this behavior is expected to be opposite.

 \section{\label{sec:level4}Conclusion}

	\begin{figure}
		\centerline{\includegraphics[width=\textwidth]{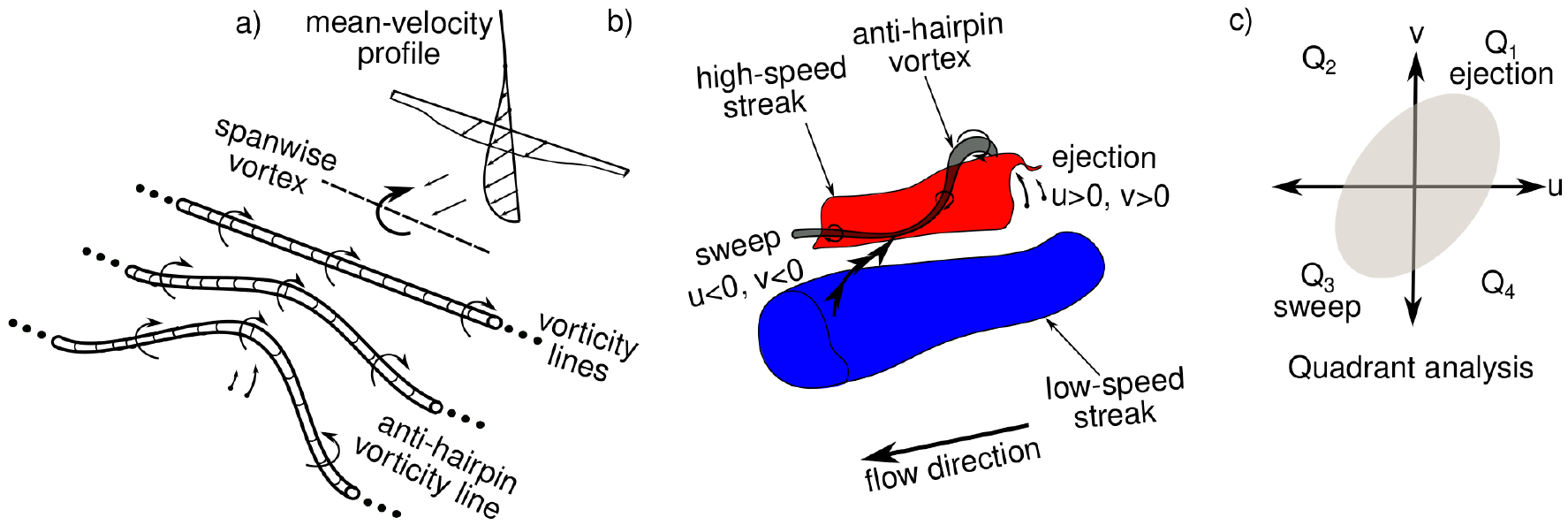}}
		
		\caption{Schematic of anti-hairpin vortex theory. (\textit{a}) Conceptual illustration of the vortex filaments during the evolution of an anti-hairpin vortex. (\textit{b}) The low-speed streak and high-speed streak with an anti-hairpin vortex. (\textit{c}) Tangential velocity fluctuation (u in the streamwise direction and v is the wall-normal direction of velocity component) is mapped on the two-dimensional coordinate systems.}
		
		\label{fig:6}
	\end{figure}
 
 The typical coherent structures for buoyancy-driven turbulent boundary layers were observed anti-hairpin vortices which are pointed upstream direction. The anti-hairpin vortex is related dominant spanwise vortex located outer layer of the peak velocity in the mean velocity profile (Fig. \ref{fig:6}a). Subsequently, the spanwise vortex lifts away from the wall, pushes the fluid particle in the downstream direction (u $>$ 0 and v $>$ 0), is also related to the ejection events. Meanwhile, the ejections events feed by legs and necks of the anti-hairpin vortex from the low-speed streaks; fluid particle moves toward the wall referred to sweep events (u $<$ 0 and v $<$ 0). Not only the pattern of the vortex was V-shaped but also the characteristics were opposite to the hairpin vortex. The anti-hairpin vortex straddles on the high-speed streak and is opposite rotational direction than the hairpin vortex (Fig. \ref{fig:6}b). The most probability of u-v correlation is positive; the first quadrant related to the ejection whereas the third quadrant is related to the sweep event(Fig \ref{fig:6}c). The ejection events are observed stronger and greater speed than the sweep events, meanwhile, low-speed streaks tend to be larger than the high-speed streaks.

\begin{acknowledgments}
The numerical calculations reported in this paper were fully performed at TUBITAK ULAKBIM, High Performance and Grid Computing Center (TRUBA resources).
\end{acknowledgments}

\section*{Declaration of interests}
	
	The authors declare that they have no known competing financial interests or personal relationships that could have appeared to influence the work reported in this paper.

\nocite{*}
\bibliography{sorsamp}

\end{document}